\documentclass{aa}
\usepackage{amsmath}
\usepackage{amssymb}
\usepackage[T1]{fontenc}
\usepackage{graphicx}
\usepackage{natbib}
\usepackage{type1cm}

\bibpunct{(}{)}{;}{a}{}{,}

\begin{document}

   \title{Refined numerical models for multidimensional Type Ia supernova
          simulations}

   \author{M.\ Reinecke \and
           W.\ Hillebrandt \and
           J.C.\ Niemeyer
          }

   \offprints{M.\ Reinecke \\({\itshape martin@mpa-garching.mpg.de})}

   \institute{Max-Planck-Institut f\"ur Astrophysik,
              Karl-Schwarzschild-Str.\ 1, 85741 Garching, Germany}

   \date{Received ?? ??, 2001; accepted ?? ??, 200?}

   \abstract{
      Following up on earlier work on this topic \citep{reinecke-etal-99a,
      reinecke-etal-99b}, we present an improved set of numerical models for
      simulations of white dwarfs exploding as Type Ia supernovae (SNe Ia).
      Two-dimensional simulations were used to test the reliability
      and numerical robustness of these algorithms; the results indicate
      that integral quantities like the total energy release are
      insensitive to changes of the grid resolution (above a certain
      threshold), which was not the case for our former code.\\
      The models were further enhanced to allow fully three-dimensional
      simulations of SNe Ia. A direct comparison of a 2D and a 3D calculation
      with identical initial conditions shows that the explosion is
      considerably more energetic in three dimensions; this is most likely
      caused by the assumption of axisymmetry in 2D, which inhibits the growth
      of flame instabilities in the azimuthal direction and thereby decreases
      the flame surface.
   }
   \maketitle

\section{Introduction}
The project described in this article is a direct continuation of the work
presented by \cite{reinecke-etal-99a}, and is therefore concerned with the
accurate modeling of the rapid thermonuclear combustion processes which occur
in a white dwarf during its explosion as a Type Ia supernova.
While the earlier results already gave several new insights into the explosion
dynamics of such an event and its dependence on parameters like the ignition
conditions, it nevertheless became clear that the numerical models employed
at that time were not sufficient to reproduce characteristic properties of
SNe Ia that are known from observations; most importantly, the total energy
release predicted by the simulations lay consistently below the expectations.
Meanwhile, several possible reasons for this deficiency could be
identified and the models were enhanced and adjusted accordingly, leading to
much more realistic results.

The following section gives details of the implemented changes and explains
why they were required for the correct treatment of SNe Ia. These modifications
affect the numerical description of the flame as well as the model for
the unresolved turbulent motions and the thermonuclear reaction rates.
Section \ref{calc2d} presents the results obtained from a two-dimensional
resolution study, which is mainly intended to test the correctness and
numerical robustness of the code, even though data obtained from 2D simulations
is unlikely to agree with observations for reasons given in section
\ref{discuss}.

The first three-dimensional simulation performed with the new code is
discussed in section \ref{calc3d}, with a focus on analyzing the similarities
and differences compared to the axisymmetric 2D model.

Finally, in section \ref{discuss} our data are compared with observations and
numerical simulations performed by other authors; furthermore, several possible
future improvements of the code are presented briefly.

Forthcoming articles will focus on a more detailed analysis of three-dimensional
simulations with varying initial conditions and numerical resolution.

\section{Improvements of the numerical models}
It must be kept in mind that the corrections and additional
features introduced below aim at a more realistic description of the explosion
dynamics and energetics. No effort was made to reproduce other aspects,
like the exact chemical composition of the burned material, unless this was
required for the goals mentioned above. As a consequence, the treatment of
nuclear reactions might appear minimalistic in comparison to the large
reaction networks typically used in one-dimensional simulations; but
reaching a comparable level of detail in multidimensions would be prohibitively
expensive and is fortunately not required for reliable predictions of the
explosion strength.
\subsection{Nuclear reactions}
\label{nucreac}
    In a SN Ia, the nuclear reactions can be distinguished depending on their
    time scales:
    \begin{itemize}
    \item The ``fast'' reactions, which transform the initial C/O
      mixture into nuclear statistic equilibium (NSE) at high densities
      or into a mixture of
      intermediate-mass elements at lower densities, take place at time
      scales $\tau_{\text{fast}}$ that are much shorter than the time
      $\tau_{\text{cross}}$ during which the flame crosses a grid cell:
      \begin{equation}
        \tau_{\text{fast}}\ll \tau_{\text{cross}}=\Delta / s\text{,}
      \end{equation}
      where $\Delta$ denotes the linear extent of a cell and $s$ the flame
      propagation speed.
      Therefore the transition between unburned and burned material is too
      abrupt to be spatially resolved and should be modeled as a numerical
      discontinuity. This can be done by using the so-called \emph{level
      set technique} (see section \ref{thinflame}).
    \item Composition changes due to the expansion of the star (i.e.\ changes
      of the NSE), on the other hand, are much slower than $\tau_{\text{cross}}$
      and the Courant time step and can be modeled by simply changing the local
      species concentrations to the desired values after each time step.
    \end{itemize}
    This picture is not entirely correct, since there exist -- especially at
    lower densities -- reactions behind the flame whose time scales are
    comparable to $\tau_{\text{cross}}$. However, these reactions have no
    significant effect on the internal energy or the molecular weight of the
    material and therefore can be neglected if one is only interested in
    the hydrodynamics of the explosion.

    The original version of the combustion algorithm (see
    \citealt{reinecke-etal-99a}) did not contain any slow reactions,
    and the fast reactions were approximated rather crudely by the instant
    fusion of carbon and oxygen to nickel. In the meantime we realized that
    such a description is oversimplified and cannot lead to accurate results
    in a Type Ia model for several reasons:
    \begin{itemize}
      \item Combustion at high densities does not immediately produce
      nickel, but also a considerable fraction of $\alpha$-particles.
      The specific energy release of this reaction is much lower than for the
      production of pure nickel.
      \item However, as the density in the ashes drops due to the stellar
      bulk expansion, the equilibrium shifts towards the heavy elements
      and the energy ``stored'' in the $\alpha$-particles is released during
      the later explosion stages.
      \item As soon as the flame enters regions of intermediate density
      ($10^8$--$10^7$g/cm$^3$), elements of intermediate mass are synthesized
      instead of nickel, and the specific energy release drops accordingly.
    \end{itemize}

    Since all of these formerly neglected phenomena may affect the explosion
    dynamics, they were implemented in our new scheme, which is described
    below:

  \begin{itemize}
    \item The initial mixture consists of $^{12}$C and $^{16}$O at low
      temperatures. Because of the electron degeneracy the fuel temperature is
      nearly decoupled from the rest of the thermodynamical quantities, and
      since temperature is not used to determine the initial reaction rates,
      its exact value is unimportant.
    \item When the flame passes through the fuel, carbon and oxygen are
      converted to ash, which has different compositions depending on the
      density of the unburned material. At high densities
      ($\rho>5.25\cdot 10^7$\,g/cm$^3$),
      a mixture of $^{56}$Ni and $\alpha$-particles in nuclear statistic
      equilibrium is synthesized. Below that density burning only produces
      intermediate mass elements, which are represented by $^{24}$Mg. Once
      the density drops below $10^7$\,g/cm$^3$, no burning takes place.
    \item Because NSE is assumed in the ashes, the proportion of
      $^{56}$Ni and $\alpha$-particles will change depending on density and
      temperature. To obtain the correct values, a NSE data table provided by
      H.-Th.\ Janka and an iterative root finding algorithm were employed.
    \item Whenever the composition in a cell changes due to nuclear reactions,
      a corresponding amount of energy is added or subtracted from the total
      energy in that cell. The amount is derived from the nuclear binding
      energies published by \cite{audi-etal-97}.
  \end{itemize}
Compared to the much simpler original approach,
this more detailed and realistic model has the
potential to alter the explosion dynamics significantly: first, the reaction
to NSE and intermediate elements with its comparatively low heat release
results in lower temperature and higher density of the burned material, which
will decrease the buoyancy of the burning bubbles. In addition the number of
particles per unit mass in the ashes is also higher with the new model, which
reduces the temperature even more. Since the $\alpha$-particles will be
converted to nickel during the expansion of the star, not all of the
nuclear energy is released at once when material is processed by the flame,
but the release is partially delayed. The combination of all these effects
is expected to shift the maximum of energy generation to a later time during
the explosion.

The transition densities from burning to NSE and incomplete burning, as well as
from incomplete burning to flame extinction were derived from data of a W7 run
provided by K.\ Nomoto. This approach is rather phenomenological, and
since these densities can have a potentially large impact on the simulation
outcome, it will have to be re-examined in a thorough manner. 

\subsection {Thin flame model}
\label{thinflame}

  The numerical representation of the thermonuclear reaction front (i.e.\ the
  location where the ``fast'' reactions take place) did not change
  fundamentally compared to the original implementation
  \citep{reinecke-etal-99a}.
  To repeat briefly, the flame front is associated with the zero level
  set of a function $G(\vec r, t)$, whose temporal evolution is given by
  \begin{equation}
     \frac{\partial G}{\partial t}
      = - (\vec{v}_u+s_u\vec{n})(-\vec{n}|\vec{\nabla}G|)\text{,}
      \label{gprop}
  \end{equation}
  where $\vec{v}_u$ and $s_u$ denote the fluid and flame propagation velocity
  in the unburned material ahead of the front, and $\vec n$ is the front normal
  pointing towards the fuel.
  The advection of $G$ caused by the fluid motions is
  treated by the piecewise parabolic method \citep{colella-woodward-84},
  which is also used by our code to integrate
  the Euler equations. After each time step, the front is additionally 
  advanced by $s_u \Delta t$ normal to itself.

  This equation is only applied in the close
  vicinity of the front, whereas in the other regions $G$ is adjusted such
  that
  \begin{equation}
    |\vec \nabla G| = 1\text{.}
  \end{equation}

  The source terms for energy and composition due to the fast thermonuclear
  reactions in every grid cell are determined as follows:
  \begin{align}
    X'_{\text{Ashes}} &= \text{max}(1-\alpha, X_{\text{Ashes}}) \label{newash} \\
    X'_{\text{Fuel}} &= 1-X'_{\text{Ashes}} \\
    e'_{\text{tot}} &= e_{\text{tot}} + q (X'_{\text{Ashes}}-X_{\text{Ashes}})\text{,}
   \end{align}
  where $\alpha$ is the volume fraction of the cell occupied by unburned
  material; this quantity can be determined from the values of $G$ in the
  cell and its neighbours. The quantity $q$ represents the specific energy
  release of the total reaction.

  In the new model for the nuclear reactions (see section \ref{nucreac})
  the composition of the ashes depends on the density of the
  unburned material ahead of the front, which cannot be properly determined
  from the state variables of the grid cell in question, since it contains
  a mixture of burned and unburned states from which a reconstruction is not
  possible. Currently the maximum density of all neighbour cells is adopted
  as an approximation for the real density of the unburned material and used
  to determine the compositon of the newly created ashes. This approach is
  acceptable in the context of SNe Ia, since the white dwarf does not contain
  steep density gradients (with the exception of the flame itself), and
  supersonic phenomena, which might create such gradients, do not occur.

\subsection{Model for the turbulent flame speed}
  All multidimensional simulations of exploding white dwarfs share the
  problem that it is impossible to resolve all hydrodynamically unstable
  scales. The consequence is that the simulated thermonuclear flame can
  only develop structures on the resolved macroscopic scales, while the real
  reaction front will be folded and wrinkled on scales down to the Gibson scale
  $l_g$, which is indirectly defined by
  \begin{equation}
    v'(l_g) = s_l\text{,}
  \end{equation}
  where $v'(l)$ denotes the amplitude of velocity fluctuations on a scale $l$
  and $s_l$ the laminar flame propagation speed. In our simulations, $l_g$
  almost always lies far below the grid scale. Simply neglecting the surface
  increase
  on sub-grid scales would lead to an underestimation of the energy generation
  rate, which is not acceptable; therefore a model for a turbulent flame speed
  $s_t > s_l$ is required to compensate this effect.

  For the case of very strong turbulence (i.e.\ for $v'(\Delta) \gg s_l$ or
  $l_g \ll \Delta$ for a numerical resolution of $\Delta$) it has been shown
  that the turbulent flame velocity
  decouples from $s_l$ and is proportional to the velocity fluctuations $v'$
  \citep{shchelkin-43,pocheau-92,peters-00}. In our simulations this condition
  is fulfilled except at the very beginning of the explosion, where combustion
  starts laminar before sufficient turbulence has been generated; this stage
  only lasts a few tens of milliseconds. At lower densities ($\rho \lessapprox
  10^7$g/cm$^3$) this picture is possibly not correct, since in that
  case the combustion takes place in the so-called \emph{distributed burning
  regime} \citep{peters-86}; it is still a matter of debate whether a transition
  to a detonation can occur at these densities
  \citep{niemeyer-99,khokhlov-etal-97},
  which would of course lead to a quite different
  chemical composition and a higher energy release in the outer stellar layers.
  For the time being we ignore the possibility of a delayed detonation; if
  further work in this field indicates that this scenario cannot be ruled out,
  however, this point will have to be addressed.

  As discussed above, the quantity $v'(\Delta)$ must be known in order to
  determine the turbulent macroscopic flame speed $s_t$. Unfortunately it cannot
  be determined directly from the Reynolds stress tensor at the location of the
  front because of the velocity jump between burned and unburned material.
  Therefore we make use of a technique first presented by \cite{clement-93} and
  later applied to SNe Ia by \cite{niemeyer-hillebrandt-95a}, which models
  the creation, advection and dissipation of the turbulent kinetic energy on
  sub-grid scales $q$ from which $v'(\Delta)$ can be easily derived.

  The time evolution of $q$ is given by
    \begin{equation}
      \frac{\partial (\rho q)}{\partial t}
      + \vec\nabla(\bar{\vec v}\rho q) =
      -\frac{2}{3} \rho q \vec\nabla \bar{\vec v}
      + \Sigma_{ij}\frac{\partial \bar v_i}{\partial x_j}
      - \rho e_{\text{diss}}+\vec\nabla(\eta_{\text{turb}}\vec\nabla q)\text{.}
      \label{qtime}
    \end{equation}
The individual source and sink terms on the right hand side can be interpreted
as turbulent compression, input from macroscopic scales, dissipation into
thermal energy and turbulent diffusion, respectively.

Following \cite{clement-93}, the turbulent viscosity, stress tensor and energy
dissipation are modeled by
    \begin{equation}
      \label{etaturb}
      \eta_{\text{turb}} =\rho \Delta v' (\Delta)=\rho\mathcal{C}\Delta\sqrt{q}
      \text{,}
    \end{equation}
    \begin{equation}
    \Sigma_{ij}=\eta_{\text{turb}} \left(
        \frac{\partial \bar v_i}{\partial x_j}
      + \frac{\partial \bar v_j}{\partial x_i}
      - \frac{2}{3}\delta_{ij}(\vec\nabla \bar{\vec v})\right)\text{\quad and}
      \label{sigmaturb}
    \end{equation}
    \begin{equation}
      \label{ediss}
      e_{\text{diss}} ={v'}^3 \Delta^{-1} = \mathcal{D}q^{3/2}\Delta^{-1}.
    \end{equation}
  The remaining free parameters $\mathcal{C}$ and $\mathcal{D}$ are set to
    \begin{align}
      \mathcal{C} &= 0.1 \mathcal{F} \quad \text{and}\\
      \mathcal{D} &= 0.5 \mathcal{F}^{-1}\text{,}
      \\ \text{where} \quad
      \mathcal{F} &= \text{min}(100,\text{max}(0.1, 10^{-4} e_i/q))\text{,}
    \end{align}
  again in accordance to Clement's work.

From the appearance of only one cell dimension $\Delta$ in the above equations
it is obvious that this approach is only valid for grids with quadratic and
cubic cells. For anisotropic grids, Clement suggests to calculate
$\eta_{\text{turb}}$ separately for each spatial direction, but to treat
the dissipation still as an isotropic effect. The expressions for
$\eta_{\text{turb}}$ and $e_{\text{diss}}$ then become
    \begin{align}
    \label{aniso1}
      \eta_{\text{turb},i}
        &=\rho\mathcal{C}\Delta_i^2 \thinspace \bar\Delta^{-1}\sqrt{q}\quad\text{and} \\
      e_{\text{diss}} &= \mathcal{D}q^{3/2} \thinspace \bar\Delta^{-1}\text{,}
    \label{aniso2}
    \end{align}
where $\bar\Delta$ is the geometric mean of the individual cell dimensions.
However, even if this extension is used, the model is expected to perform
rather poorly in the extreme case of very elongated cells, since the concept
of a ``cell size'', on which $q$ is defined, breaks down under these
circumstances.

The model described above produces an approximation of the turbulent sub-grid
energy throughout the computational domain, although the source and sink terms
for $q$ must be disabled in the cells cut by the front because of the velocity
jump. This is possible because $q$ is transported into these cells by means
of advection and turbulent diffusion. The desired magnitude of the velocity
fluctuations is then given by
\begin{equation}
  v'(\Delta)=\sqrt{2q},
\end{equation}
which completes the model for the flame propagation speed.

In the given form, the sub-grid model is correct for three-dimensional
simulations only; in two spatial dimensions it is necessary to replace the
factor $2/3$ in equations (\ref{qtime}) and (\ref{sigmaturb}) by a factor 1 to
preserve the tracelessness of $\Sigma_{ij}$. Leaving this factor unchanged
would imply three-dimensional turbulent motions on unresolved scales, whereas
the flow on the macroscopic scales is only two-dimensional. Since the
transition scale between both regions is identical to the resolution,
one would expect different simulation results for identical initial conditions,
if the cell sizes are varied; this is certainly not desirable.
In the older versions of the simulation code this fact was overlooked, and
all two-dimensional calculations were performed with the incorrect factor
$2/3$; this resulted in a rather strong dependence of the total energy release
from the numerical resolution. With the corrected model, however, the
resolution does not influence the explosion energetics significantly
(see section \ref{calc2d}).

It must be emphasized at this point that any attempt to simulate turbulent
flow in two dimensions will likely produce results that differ significantly
from those of fully three-dimensional simulations; this effect is caused
by the unrealistic absence of turbulent velocity fluctuations along the third
coordinate
axis, which may have a strong influence on the scaling behaviour of the
turbulent motions. Therefore the results of two-dimensional calculations
involving turbulent flow must be checked at least punctually against their
three-dimensional equivalents in order to give them the required credibility.

\section{Two-dimensional calculations}
\label{calc2d}
The goal of the calculations presented here was to verify that variations of
parameters without direct physical relevance (e.g.\ numerical resolution) do
not affect the results significantly. Because of the enormous
computing and storage requirements for this kind of numerical study, the
simulations had to be performed in two dimensions assuming axisymmetry instead
of solving the full three-dimensional problem. Nevertheless, the insights
gained from these experiments can -- at least in a qualitative manner -- also
be applied to the three-dimensional case.

In analogy to the calculations described in \cite{reinecke-etal-99b}, the
experiments were carried out in cylindrical ($r$, $z$) coordinates with a
uniform grid spacing in the inner regions of the computational domain (from
the origin to a radius of $\approx 2.3\cdot 10^8$cm). In the outer regions
the radial and axial dimensions of the grid cells were increased exponentially
in order to avoid mass loss across the grid borders due to the stellar
expansion. Equatorial symmetry was assumed; at the same time rotational
symmetry around the polar axis was imposed by the choice of coordinates.
The white dwarf was constructed by assuming a central density of
$2.9\cdot 10^9$g/cm$^3$, a uniform temperature of $5\cdot10^5$K and a
composition of $X_C=X_O=0.5$, resulting in a mass of $2.797\cdot10^{33}$g,
a radius of about $1.8\cdot10^8$cm and a binding energy of
$5.19\cdot10^{50}$erg. 

\subsection{Resolution study}

To study the robustness of our code with respect to a change of the numerical
resolution, simulations
were performed with grid sizes of $128^2$, $256^2$, $512^2$ and $1024^2$
cells, whose corresponding resolutions in the uniform inner part of the grid
were $2\cdot10^6$\,cm, $10^6$\,cm, $5\cdot10^5$\,cm and $2.5\cdot10^5$\,cm.
The initial flame geometry (called c3\_2d) used for all these calculations
is identical to
the setup C3 presented by \cite{reinecke-etal-99b}: the matter
within a radius of $1.5\cdot10^7$cm from the stellar center was incinerated,
and the surface of the burned region was perturbed to accelerate the
development of Rayleigh-Taylor instabilities. 

  \begin{figure}[tbp]
    \centerline{\includegraphics[width=0.5\textwidth]{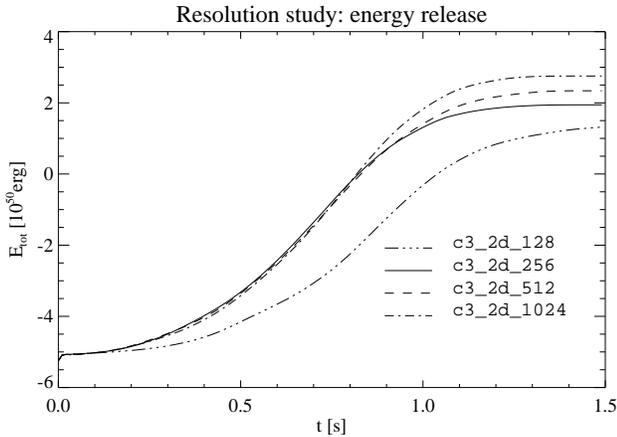}}
    \caption{Time evolution of the total energy for the initial flame geometry
             c3\_2d and different resolutions.
    During the early and intermediate explosion stages there is excellent
    agreement between the better resolved simulations.}
    \label{e_compare}
  \end{figure}
  \begin{figure*}[tbp]
    \centerline{\includegraphics[width=0.7\textwidth]{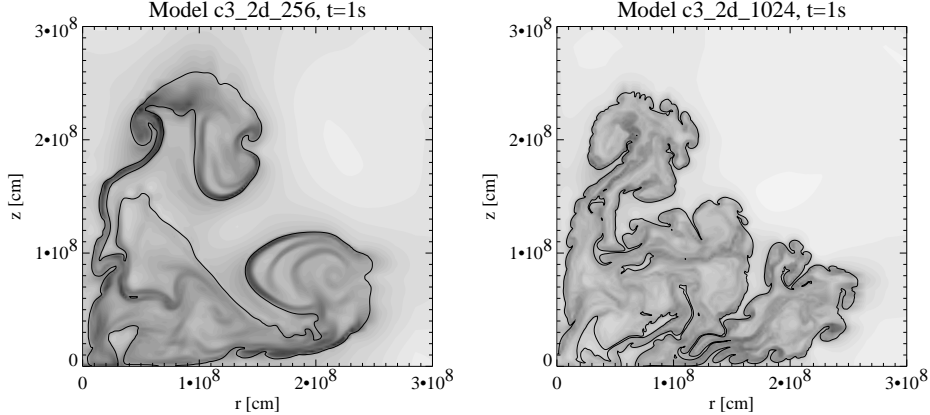}}
    \caption{Comparison of two simulations with identical initial conditions,
    but different resolution. The front geometry is indicated by a black
    line, whereas the amplitude of turbulent velocity fluctuations is given
    color-coded from 0 (white) to 2.5$\cdot$10$^7$\,cm/s (black).
    The low-resolution run clearly exhibits less flame structure, which is
    compensated by a higher flame speed. The overall features appear to be
    remarkably similar in both diagrams.}
    \label{res_compare}
  \end{figure*}
Figure \ref{e_compare} shows the energy release of the models; except for
the run with the lowest resolution, the curves are nearly identical in the
early and intermediate explosion stages. Simulation c3\_2d\_128 exhibits
a very slow initial energy increase and does not reach the same final level
as the other models. Most likely this is due to insufficient resolution,
which leads to a very coarsely discretized initial front geometry and thereby
to an underestimation of the flame surface. From this result it can be deduced
that all supernova simulations performed with our code should have a central
resolution of $10^6$\,cm or better.

  In the late explosion phase (after about 0.8 seconds) the total energy
  is no longer equal for the three simulations, but increases with better
  resolution.
  The origin of this effect is not clear, but since it occurs at the time
  when the flame enters the anisotropic part of the grid, it is probably
  caused by inaccuracies of the burning and hydrodynamic algorithms in the
  highly elongated cells in these regions. A breakdown of the sub-grid
  turbulence model is the most likely culprit, since in a very anisotropic
  grid the concept of a ``grid scale'' is not well defined anymore. 
  In any case the observed scatter in the final energy releases of the order of
  10\% was assumed to be acceptable.

Overall, our model for the turbulent flame speed appears to compensate the lack
of small structures in the front very well.

  Another argument for the reliability of the turbulence model is provided by
  figure \ref{res_compare}, which shows the front geometry
  and turbulence intensity after one second for the simulations c3\_2d\_256
  and c3\_2d\_1024. These data are in very good agreement with the expectations:
  \begin{itemize}
    \item Both simulations have quite similar large-scale features;
      e.g.\ the burned
      areas appear similar in size, and the largest distance from the front
      to the center also is nearly equal.
    \item Because of the larger scale range in c3\_2d\_1024, the flame
      structure is much better resolved: it shows, for example, the onset of
      Kelvin-Helmholtz-instabilities in the shear flow between rising hot
      material and falling fuel near the coordinate axes and also small,
      secondary Rayleigh-Taylor instabilities, which are not
      visible in the coarser simulation.
    \item The amplitude of sub-grid fluctuations is lower for the better resolved
      simulation, which is in agreement with the turbulent velocity scaling law.
      This is also required to compensate the larger flame surface and keep
      the total energy generation rate resolution-independent.
  \end{itemize}
  This plot also demonstrates convincingly that the highest turbulence
  intensities are reached in the shear layer between fuel and ashes.

\subsection{Comparison to earlier 2D simulations}

Comparison of the energy evolution with the results presented by
\cite{reinecke-etal-99b} reveals that the final absolute energies are
considerably higher for the refined code than for the older calculations.
The remnant, which had formerly remained bound, now reaches a positive total
energy and will therefore continue its expansion. However, the energy release
is still too low to account for a typical SN Ia.

\begin{figure}[tbp]
    \centerline{\includegraphics[width=0.5\textwidth]{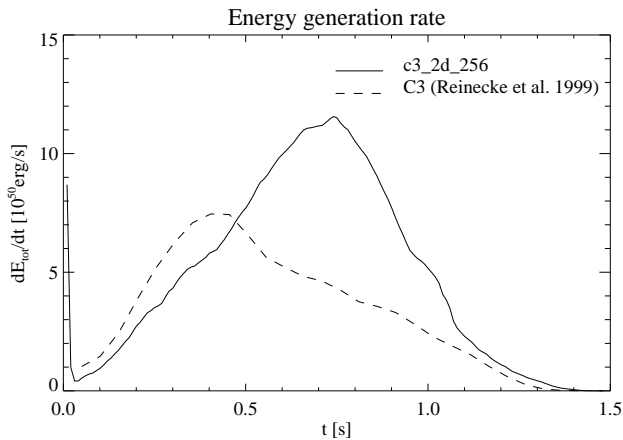}}
    \caption{Time evolution of the energy generation rate for the models
       C3 in \cite{reinecke-etal-99b} and c3\_2d\_265. The shift of the
       maximum is clearly visible.}
    \label{eprod}
  \end{figure}

Significant differences also exist in the time evolution of the energy
generation rate (figure \ref{eprod}):
the period of most intense burning, which was originally reached after
0.2\,--\,0.5\,s, has now shifted to 0.5\,--\,0.8\,s after ignition.
The main reason for this delayed energy release is the more realistic treatment
of the fusion reactions. Since carbon and oxygen are no longer instantaneously
fused to nickel, but to a NSE mixture of nickel and $\alpha$-particles, less
energy is released in the early burning phases, which results in a lower
temperature and higher density in the ashes. This in turn decreases the
buoyancy and rising speed of the burning bubbles, leading to less shear
and lower turbulent burning speeds. As a compensation, the transformation
of $\alpha$-particles into nickel injects additional energy into the explosion
at later times.

\begin{figure}[tbp]
    \centerline{\includegraphics[width=0.5\textwidth]{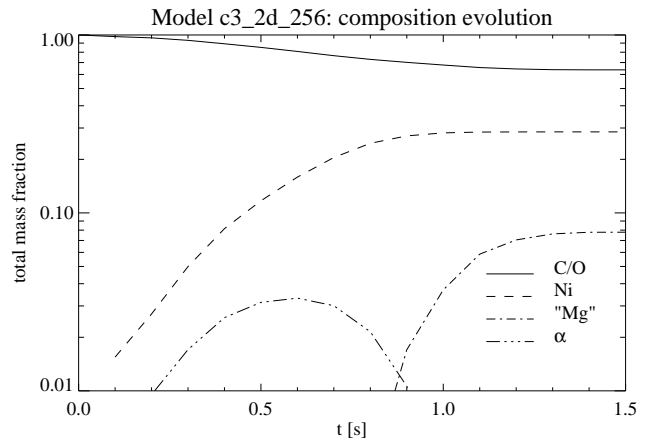}}
    \caption{Time evolution of the chemical composition for model c3\_2d\_256.}
    \label{nuc_evolve2d}
  \end{figure}

This argument is supported by an evolution diagram of the
white dwarf's composition, shown in figure \ref{nuc_evolve2d}. As long as the
nuclear reactions take place near the center and bulk expansion is still
rather slow, nickel and $\alpha$-particles are produced at a nearly constant
ratio, and the specific energy release is much lower than for synthesis of
pure nickel. During the continued expansion, the equilibrium concentration
for $\alpha$-particles in the NSE rapidly drops towards zero and the
``buffered'' energy is released as thermal energy, further driving the
expansion. This very exothermic process ($Q\approx 1.52\cdot 10^{18}$\,erg/g)
coincides rather well with the maximum of the energy generation rate
(figure \ref{eprod}).

\section{Extension to three dimensions}
\label{calc3d}
In order to compare two- and three-dimensional simulations directly, a 3D
calculation was performed using the same initial conditions as given in
section \ref{calc2d}.
For this purpose the initial two-dimensional flame location
was rotated by 90 degrees around the
$z$-axis and mapped onto the three-dimensional Cartesian grid consisting
of $256^3$ cells with a central resolution of $10^6$\,cm. Only one octant
of the white dwarf was simulated and mirror symmetry was assumed with respect
to the coordinate planes.

  \begin{figure}[tbp]
    \centerline{\includegraphics[width=0.5\textwidth]{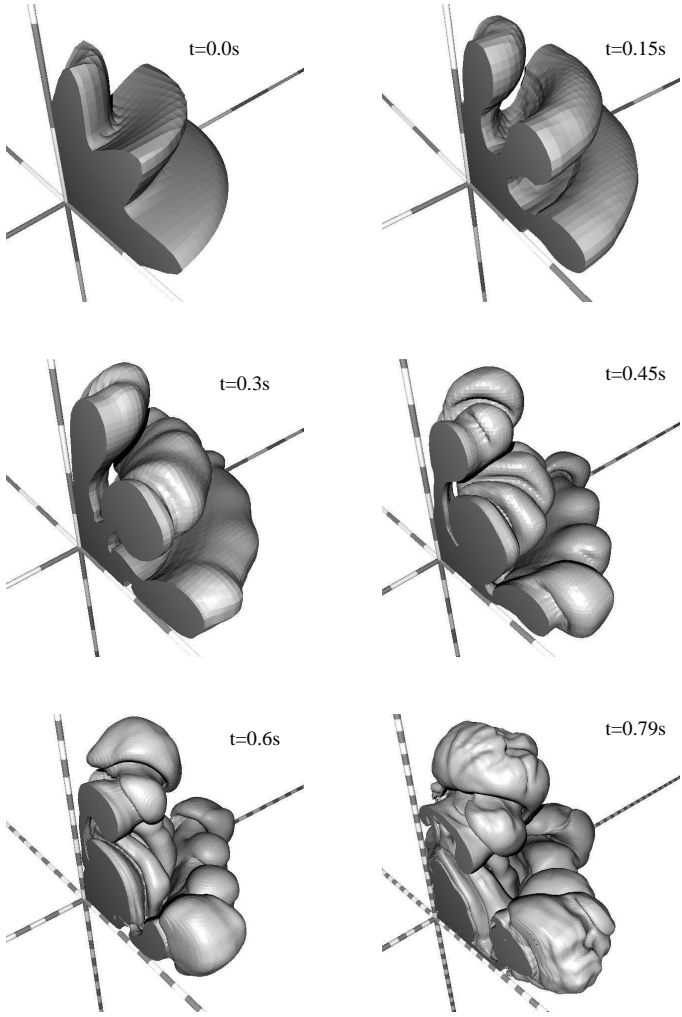}}
    \caption{Snapshots of the flame front for a centrally ignited
       three-dimensional scenario. One ring on the coordinate axes corresponds
       to $10^7$cm.}
    \label{c3_3d_front}
  \end{figure}
The initial configuration, as well as snapshots at later times, are shown
in figure \ref{c3_3d_front}. Obviously, the initial axisymmetry is lost
after 0.2\,--\,0.3\,s, although
  no explicit perturbation in $\varphi$-direction was applied to the front.
  This happens because the initial flame geometry cannot be mapped perfectly
  onto a Cartesian grid and therefore the front is not transported at exactly
  the same speed for all $\varphi$. On a cylindrical grid, which matches the
  symmetry of the problem setup better, this effect would not be observed.

  In our case, however, this symmetry breaking is desired, since an exactly
  axisymmetric calculation would only reproduce the results of the 2D
  run. Besides, perfect rotational symmetry would never occur in reality.

\begin{figure}[tbp]
    \centerline{\includegraphics[width=0.5\textwidth]{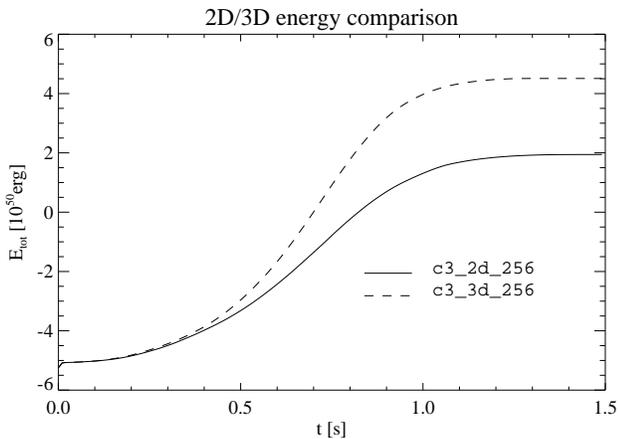}}
    \caption{Comparison of the explosion energy for identical initial conditions
     and resolution in two and three dimensions. After the loss of axial
     symmetry (at $t\approx0.3$\,s) the larger flame surface in the
     three-dimensional model leads to more vigorous burning.}
    \label{comp2d3d}
  \end{figure}

  During the next few tenths of a second, the small deviations cause the
  formation of fully three-dimensional RT-mushrooms, leading to a
  strong convolution of the flame. As expected, this phenomenon
  has a noticeable influence on the explosion energetics; this is illustrated
  in figure \ref{comp2d3d}.
  Before the loss of axial symmetry in c3\_3d\_256, the total energy evolution
  is almost identical for both simulations, which strongly suggests that the
  two- and three-dimensional forms of the employed turbulence and level set
  models are consistent, i.e.\ that no errors were introduced during the
  extension of these models to three dimensions. In the later phases the 3D
  model releases more energy as a direct consequence of the surface increase
  shown in figure \ref{c3_3d_front}.

\section{Discussion and conclusions}
\label{discuss}

Since the explosion energy and the composition of the remnant are important
criteria for the quality of a SN Ia model, the relevant data for all
performed simulations are listed in table \ref{burntable}. 

\begin{table}[htbp]
  \centerline{
    \begin{tabular}{|l|c|c|c|}
    \hline
    model name & $\frac{m_{\text{Mg}}}{M_\odot}$&$\frac{m_{\text{Ni}}}{M_\odot}$&$\frac{E_{\text{nuc}}}{10^{50}\text{\,erg}}$\vphantom{\raisebox{2pt}{\large A}}\vphantom{\raisebox{-7pt}{\large A}} \\
    \hline
    c3\_2d\_128 & 0.132 & 0.348 & 6.56 \\ \hline
    c3\_2d\_256 & 0.109 & 0.400 & 7.19 \\ \hline
    c3\_2d\_512 & 0.151 & 0.402 & 7.58 \\ \hline
    c3\_2d\_1024 & 0.152 & 0.428 & 8.00 \\ \hline
    c3\_3d\_256 & 0.177 & 0.526 & 9.76 \\ \hline
    \end{tabular}
  }
  \caption{Overview over element production and energy release of all
    discussed supernova simulations}
  \label{burntable}
\end{table}

Given the stellar binding energy of $5.19\cdot 10^{50}$\,erg, it is evident
that the progenitor becomes unbound in all experiments, which implies that no
recontraction will occur and no compact object remains. Nevertheless only
model c3\_3d\_256 results in a powerful enough explosion to qualify as a
typical SN Ia; the two-dimensional scenarios are too weak to accelerate
the ejecta to the speeds observed in real events and produce too little nickel
to power a standard SN Ia light curve, which had more or less to be expected,
since the surface growth was partially suppressed.

The 3D calculation is a good candidate
for typical SN Ia explosions, at least with respect to explosion strength and
remnant composition. Its nickel mass falls well into the range of
$\approx$\,0.45\,--\,0.7\,$M_{\odot}$ determined by \cite{contardo-etal-00} for several typical events,
and it can be deduced from the amount of ``magnesium'' in the ejecta that
enough intermediate mass elements were synthesized to explain the observed
spectral features.

Qualitatively, our results for the explosion energetics are in rather good
agreement with recent simulations performed by \cite{khokhlov-00}, which
rely on quite different numerical models and assumptions; most notably
the macroscopic flame velocity is determined by the asymptotic rise speed of a
Rayleigh-Taylor bubble with size $\Delta$, instead of the turbulent velocity
fluctuations.
The particular setup described in his paper is centrally ignited, and no
perturbation is applied to the spherical flame surface. Due to the high
resolution the discretization errors are quite small, and there is a rather
long phase of very slow burning until the first instabilities have reached
the nonlinear stage; from this moment on, however, the explosion progresses
in a way very similar to the 3D simulation discussed here.
This similarity applies to the energy production rate as well as the
geometrical features developed by the flame.

Further improvements of the SN Ia simulations are planned, both with respect
to the underlying models and the realism of the initial conditions.
As was pointed out by S.~Blinnikov (personal communication), an accurate
simulation of light curves and spectra based on the results of the 3D
hydrodynamical computation requires not only the total concentration of iron
group elements (which have been represented so far by $^{56}$Ni), but the
amounts of the real $^{56}$Ni and the rest of the Fe-group elements.
The necessary changes will be implemented in the near future.

It would also be desirable to reach a higher numerical resolution at the
beginning of the explosion, which allows a more accurate prescription of the
flame geometry, and, at the same time, to follow the expansion
of the star over a longer time period (several seconds), until the ballistic
expansion stage has been reached. To achieve both of these goals, an expanding
grid must be used instead of the currently employed static one.

\begin{acknowledgements}
The authors thank Ken'ichi Nomoto for pointing out the necessity to incorporate
the NSE effects into our burning model and for providing W7 datasets, and
Hans-Thomas Janka, who provided the required NSE data tables.

This work was supported in part by the Deutsche Forschungsgemeinschaft under
Grant Hi 534/3-3.
The numerical computations were carried out on a Hitachi SR-8000 at the
Leibniz-Rechenzentrum M\"unchen as a part of the project H007Z.
\end{acknowledgements}

\bibliographystyle{apj}
\bibliography{refs}

\end{document}